\documentclass[sigconf,natbib=false]{acmart}

\usepackage{hyperref}
\usepackage{algpseudocode}
\usepackage{algorithm}
\usepackage{tikz}
\usepackage{subcaption}
\usepackage[sorting=none]{biblatex}

\acmConference[SC'25]{The International Conference for High Performance Computing, Networking, Storage, and Analysis}{November 16--21}{St. Louis, MO, USA}

\title{Scaling the memory wall using mixed-precision - HPG-MxP on an exascale machine}

\author{Aditya Kashi}
\email{kashia@ornl.gov}
\affiliation{%
\institution{Oak Ridge National Laboratory}
\department{National Center for Computational Sciences}
\city{Oak Ridge}
\state{TN}
\country{USA}
}

\author{Nicholson Koukpaizan}
\email{koukpaizannk@ornl.gov}
\affiliation{%
\institution{Oak Ridge National Laboratory}
\department{National Center for Computational Sciences}
\city{Oak Ridge}
\state{TN}
\country{USA}
}
\author{Hao Lu}
\email{luh1@ornl.gov}
\affiliation{%
\institution{Oak Ridge National Laboratory}
\department{National Center for Computational Sciences}
\city{Oak Ridge}
\state{TN}
\country{USA}
}
\author{Michael Matheson}
\email{mathesonma@ornl.gov}
\affiliation{%
\institution{Oak Ridge National Laboratory}
\department{National Center for Computational Sciences}
\city{Oak Ridge}
\state{TN}
\country{USA}
}
\author{Sarp Oral}
\email{oralhs@ornl.gov}
\affiliation{%
\institution{Oak Ridge National Laboratory}
\department{National Center for Computational Sciences}
\city{Oak Ridge}
\state{TN}
\country{USA}
}
\author{Feiyi Wang}
\email{fwang2@ornl.gov}
\affiliation{%
\institution{Oak Ridge National Laboratory}
\department{National Center for Computational Sciences}
\city{Oak Ridge}
\state{TN}
\country{USA}
}

\date{March 2025}

\let\bld\boldsymbol

\begin{abstract}

Mixed-precision algorithms have been proposed as a way for scientific computing to benefit from some of the gains seen for artificial intelligence (AI) on recent high performance computing (HPC) platforms. A few applications dominated by dense matrix operations have seen substantial speedups by utilizing low precision formats such as FP16. However, a majority of scientific simulation applications are memory bandwidth limited. Beyond preliminary studies, the practical gain from using mixed-precision algorithms on a given HPC system is largely unclear.

The High Performance GMRES Mixed Precision (HPG-MxP) benchmark has been proposed to measure the useful performance of a HPC system on sparse matrix-based mixed-precision applications. In this work, we present a highly optimized implementation of the HPG-MxP benchmark for an exascale system and describe our algorithm enhancements. We show for the first time a speedup of 1.6$\times$ using a combination of double- and single-precision on modern GPU-based supercomputers.
\end{abstract}

\begin{document}

\maketitle

\section{Introduction}
Scientific computing has long relied on techniques from numerical linear algebra, nonlinear equations, numerical optimization, and ordinary and partial differential equations (PDEs) to model physical phenomena, make predictions, and explain them.
Even in the age of artificial intelligence (AI), these continue to remain important \cite{deelman_hpc_crossroads_2025}.

Meanwhile, driven by the insatiable arithmetic compute throughput needs of large language models, vendors of high-performance computing (HPC) hardware are designing chips increasingly geared towards extremely high performance in dense matrix-matrix multiplication (GEMM) kernels in low precision formats.
Graphics processing unit (GPU) vendors NVIDIA and AMD have designed `tensor cores' and `matrix units' respectively for a higher rate of growth in FP16, BF16 and INT8 GEMM throughput compared to IEEE FP32 and FP64 formats typically favored by scientific computing.
Indeed, NVIDIA has championed support for FP8, FP6 and FP4 formats and claimed hundreds of petaflops on a single chip \cite{nv_blackwell}, unthinkable a few years ago. The energy usage per operation is also lower when tested on matrix-matrix multiplication workloads.
In fact, energy savings from mixing the use of lower precision formats has been shown in the past even for other non-AI workloads \cite{anzt_cfd_energy_2010,azzam_half_dense_energy_2018}.

However, since many scientific computing applications rely on higher precision, typically IEEE FP64, this throughput and efficiency is difficult to access for scientific computing.
Computational scientists are thus starting to look into the use of lower precision formats in mixed-precision numerical algorithms \cite{mxp_review_2025}.

Furthermore, many real-world scientific and engineering computing workloads are limited by memory bandwidth rather than arithmetic throughput \cite{petsc_fun3d_2001}.
These include a large number of simulation applications that rely on solving PDEs governing fluid mechanics, solid mechanics, heat transfer, electromagnetism, chemically-reacting flows and plasma physics using methods such as finite differences, finite volumes, finite elements and lattice Boltzmann.
Such applications do not use tensor cores at all since they do not rely on GEMM as their main computational motif.
The primary computational motifs in simulation codes tend to be from sparse linear algebra - sparse matrix vector products (SpMV), sparse triangular solves (SpTRSV), sparse matrix sparse matrix products (SpGEMM), and dot products (DOT).
At a slightly higher level, multigrid methods play an important role in the scalable solution of PDEs, and present unique challenges in accelerated distributed HPC \cite{AMG_road_exa_2012}.
The US Exascale Computing Project invested in developing mixed precision numerical algorithms for some of these motifs \cite{ecp_mxp_advances_2021}.

The High Performance GMRES Mixed Precision (HPG-MxP) benchmark was proposed \cite{hpgmp} to measure the performance of a supercomputer on such memory-bandwidth limited simulation workloads. In contrast to the existing High Performance Conjugate Gradient benchmark \cite{dongarra_hpcg_2016}, it allows the use of mixed precision internally, while requiring a solution `somewhat close' to that obtained by a fully double-precision solver (to be clarified later).
In our view, the objective of the benchmark is to get a practical upper limit for the performance of mixed-precision memory-bandwidth limited workloads while achieving essentially the same usefulness as double-precision computation.
The work of developing and running at scale an optimized benchmark achieving the maximum possible performance will serve several purposes:
\begin{enumerate}
    \item it will serve as a yardstick to shoot for while optimizing the performance of scientific simulations applications, especially ones utilizing implicit solvers that require the solution of large sparse linear systems, 
    \item the learning from this activity will help guide computational scientists and HPC engineers on the best ways to utilize mixed-precision methods to accelerate workloads on their HPC systems, and
    \item it will guide hardware vendors and other library providers in designing and optimizing their numerical libraries to best support mixed-precision simulation workloads.
\end{enumerate}

Our contributions in this paper are as follows.
\begin{enumerate}
    \item We describe a state-of-the-art implementation of the HPG-MxP benchmark that achieves much higher performance than the reference implementation on large-scale GPU-based HPC systems.
    \item We show that with such an optimized implementation, a higher speedup than earlier reported is possible from a double-single mixed-precision solver on current-generation exascale systems. While the benchmark allows for the use of any precision format in most steps of the solver algorithm, we focus on the use of single precision as the only low precision format for this paper.
    \item We report, for the first time, a full-system HPG-MxP run (9408 nodes) on the world's first exascale system, Frontier, at the Oak Ridge National Laboratory, USA, using our optimized codebase.
    \item We report some performance analysis of the benchmark code, particularly traces showing the achieved compute-communication overlap, the achieved memory bandwidth and performance relative to the roofline.
    \item  We confirm that validation on a small fixed problem size (on a single node) is sufficient to accurately penalize our mixed precision solver. This is achieved by introducing a full-scale validation that uses all available nodes and the full problem size.
\end{enumerate}

\section{Background}

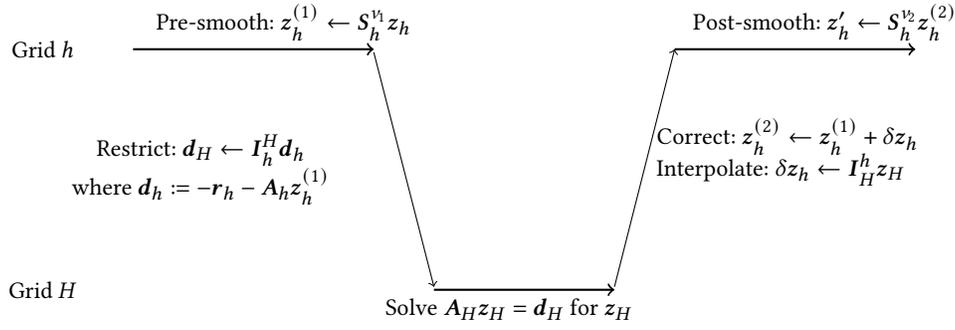
\begin{figure*}[h]
	\def\hh{4}% length of fine
	\def\Hl{3}% length of coarse
	\begin{tikzpicture}[scale=0.8] %//[node distance = 2.0cm, auto]
	\draw (-1.5,5) node[align=right] {Grid $h$};
	\draw (-1.5,1) node[align=right] {Grid $H$};
	
	\draw [->,thick] (0,5) node[align=right,xshift=2cm,above] {Pre-smooth: $\bld{z}_h^{(1)} \gets \bld{S}_h^{\nu_1}\bld{z}_h$} -- (\hh,5);
	%\node (presmooth) at (0,6) {};
	
	\draw [->] (\hh,5-0.05) -- (\hh+1,1+0.05);
	\draw (\hh+0.25,5-2) node[align=center,xshift=-2.5cm] {Restrict: $\bld{d}_H \gets \bld{I}_h^H\bld{d}_h$ \\ where $\bld{d}_h := -\bld{r}_h-\bld{A}_h\bld{z}_h^{(1)}$};
	
	\draw [->,thick] (\hh+1,1) node[align=right,xshift=1cm,below] {Solve $\bld{A}_H\bld{z}_H = \bld{d}_H$ for $\bld{z}_H$}
	-- (\hh+1+\Hl,1);
	
	\draw [->] (\hh+1+\Hl,1+0.05) -- (\hh+2+\Hl,5);
	\draw (\hh+1.25+\Hl,5-2) node[xshift=0.25cm,right] {Interpolate: $\delta\bld{z}_h \gets \bld{I}_H^h\bld{z}_H$};
	
	\draw [->,thick] (\hh+2+\Hl,4.5) node[below,xshift=1.5cm,yshift=-0.4cm] {Correct: $\bld{z}_h^{(2)} \gets \bld{z}_h^{(1)} + \delta\bld{z}_h$} (\hh+2+\Hl,5) node[align=right,xshift=2cm,above] {Post-smooth: $\bld{z}_h' \gets \bld{S}_h^{\nu_2}\bld{z}_h^{(2)}$} -- (2*\hh+2+\Hl,5);
	\end{tikzpicture}
 
	\caption{Cycle of the linear two-grid method for the system $\bld{A}_h\bld{z}_h = \bld{r}_h$. $\bld{z}_h$ is an initial guess or prior approximation of the solution on the fine grid, $\bld{S}$  is the smoothing iteration or smoother, $\bld{I}_h^H$ denotes the restriction operator, $\bld{I}_H^h$ denotes the prolongation or interpolation operator, and $\bld{z}_h'$ is the improved solution approximation.}
	\label{fig:nlinmg_2grid}
\end{figure*}

The first major attempt to make system benchmarking better reflect real scientific workloads came with the introduction of the High Performance Conjugate Gradient (HPCG) benchmark \cite{dongarra_hpcg_2016}.
Since HPG-MxP is based on HPCG, we first give an overview of HPCG.
It solves the three-dimensional Poisson equation, a fundamental PDE from which most PDE theory and solver techniques derive, using a 27-point finite difference discretization on a uniform Cartesian mesh of a cube-shaped domain.
This results in a matrix with as many rows as mesh points, and 27 non-zero values per row for interior points. Boundary points have fewer non-zeros depending on whether they lie on a face, edge or corner point.
\begin{algorithm}
    \begin{algorithmic}
        \Require Initial guess $\bld{x}_0$
        \State (preconditioner generation) $\bld{M} \gets \mathcal{P}(\bld{A})$
        \State $\bld{p}_0 \gets \bld{x}_0$, $\bld{r}_0 \gets \bld{b}-\bld{Ap}_0, \,i \gets 1$.
        \While {$i < N$}
            \State (preconditioner application) $\bld{z}_i \gets \bld{M}^{-1}\bld{r}_{i-1}$
            \If{$i = 1$}
                \State $\bld{p}_i \gets \bld{z}_i$
                \State $\alpha_i \gets (\bld{r}_{i-1}, \bld{z}_i)$
            \Else
                \State $\alpha_i \gets (\bld{r}_{i-1}, \bld{z}_i)$
                \State $\beta_i \gets \frac{\alpha_i}{\alpha_{i-1}}$
                \State $\bld{p}_i \gets \beta_i \bld{p}_{i-1} + \bld{z}_i$
            \EndIf
            \State $\alpha_i \gets (\bld{r}_{i-1}, \bld{z}_i) / (\bld{p}_i, \bld{Ap}_i)$
            \State $\bld{x}_{i+1} \gets \bld{x}_i + \alpha_i\bld{p}_i$
            \State $\bld{r}_{i} \gets \bld{r}_{i-1} - \alpha_i\bld{Ap}_i$
        \EndWhile
    \end{algorithmic}
    \caption{Preconditioned conjugate gradient algorithm for a symmetric positive-definite linear system $\bld{Ax}=\bld{b}$}
    \label{alg:pcg}
\end{algorithm}
The preconditioned Conjugate Gradient (CG) method is used as the solver. This is a Krylov subspace solver that is guaranteed to converge, for symmetric positive-definite matrices, in $N$ iterations where $N$ is the dimension of the matrix.
The version used by the benchmark \cite{dongarra_hpcg_2016} is shown in algorithm \ref{alg:pcg}.
In HPCG, the preconditioner is required to be one cycle of geometric multigrid. Multigrid is a family of highly scalable algorithms to solve a PDE-based problem \cite{brandt_multilevel_1977}, based on the ideas of error smoothing and coarse grid correction.
Multigrid methods are based on discretizing the problem on a hierarchy of meshes of coarser resolutions and applying an iterative method at each mesh, or multigrid level, to `smooth' the errors.
A two-grid cycle is shown in figure \ref{fig:nlinmg_2grid}, which is applied recursively to solve the coarse-grid problem $\bld{A}_H$ to generate a multigrid cycle.

In a well-tuned mathematically-sound implementation, for a stencil-based matrix of size $N\times N$, multigrid converges in $\mathcal{O}(N)$ operations , with a small factor \cite{brandt_multilevel_1977} (`textbook multigrid').
However, this assumes a fixed coarse grid problem size independent of $N$ on the coarsest level, with the number of multigrid levels increasing to scale up to ever larger fine-grid problem sizes.
Since HPCG fixes the number of multigrid levels to 4, this ideal scalability is not expected.

HPCG uses the symmetric Gauss-Seidel iteration as the smoother in its multigrid preconditioner.
If the system matrix $\bld{A}$ is split into lower triangular $\bld{L}$, upper triangular $\bld{U}$ and diagonal parts $\bld{D}$, the iteration for $\bld{Az} = \bld{r}$ can be written as
\begin{align}
    (\bld{D} + \bld{L}) \bld{y} &= \bld{r} - \bld{U}\bld{z}, \label{eqn:fgs} \\
    (\bld{D} + \bld{U}) \bld{z}^{(1)} &= \bld{r} - \bld{L}\bld{y},
\end{align}
where $\bld{y}$ is a temporary intermediate vector.
Note that this involves a lower SpTRSV and SpMV with an upper triangular matrix, followed by an upper SpTRSV and SpMV with a lower triangular matrix.
In an efficient implementation, the whole symmetric Gauss Seidel iteration can be done in two kernels.

There are two traditional approaches to finding parallelism in the otherwise sequential Gauss-Seidel iteration - level scheduling and multicoloring \cite[section~2.7.1]{async_thesis_2020}.
While a level-scheduled triangular solve preserves the original ordering of the matrix but only has a limited amount of parallelism, multicoloring methods use an independent set ordering to find independent sets expose more parallelism.
Typically, this means level-scheduled methods deliver the same preconditioning effectiveness to the CG solver though cannot utilize the GPU very effectively, while multicolored triangular solves may degrade the preconditioner quality somewhat (decreasing CG's convergence rate) but delivers good GPU utilization.

%Comm volume. Comm to compute ratio.
To map the problem to a distributed parallel computer, the framework of domain decomposition is used \cite{dd:gropp}.
The spatial domain under consideration, which is discretized by a mesh or graph, is partitioned among the processors. Each mesh cell or graph point is associated with a row of the matrix, and the nonzeros in that row denote its neighborhood (which may not be the same as the topological neighborhood in the mesh).
Thus, the matrix is distributed by row - each processor owns a block of rows and all columns of the matrix.

In HPCG, the processors are factored into a 3D grid, similar to the mesh itself. Thus a grid of size $N_x \times N_y \times N_z$ is uniformly divided amongst $p_x \times p_y \times p_z$ processors.
Assuming an isotropic grid that is a perfect cube both in points and processors, $N^3$ points are mapped to $p^3$ processors.
Each processor has $(\frac{N}{p})^3$ points . Let $n := N/p$.
In a typical finite difference or finite volume discretization, each mesh point or cell directly interacts with $\mathcal{O}(1)$ neighbors, that is, a constant number of neighbors independent of the problem size.
Thus, in a regular Cartesian mesh, each row has a fixed constant number of nonzeros, except for those corresponding to global boundary points.
Therefore, operations such as SpMV and Gauss-Seidel iterations compute on all the points in the subdomain owned by a processor, and need to communicate with the processors owning their nearest spatial neighboring subdomains.
For cubic subdomains as in HPCG, one requires $\mathcal{O}(n^3)$ operations of computations, and about $6n^2$ volume of communications with neighboring processors.
Seen another way, if $\nu$ is the problem size per processor, local compute scales as $\mathcal{O}(\nu)$ while communication volume scales as $\mathcal{O}(\nu^{2/3})$.
Thus, the communication volume is a geometric order lower than the local computation volume, and therefore, network bandwidth is typically not the limiting factor for HPCG performance.
In most cases, HPCG is limited by local memory bandwidth due to the low constant arithmetic intensity of the computations.

HPCG is a good measure of a HPC system's performance on real workloads that are limited by memory bandwidth.
However, many real-world problems involve nonsymmetric matrices, to which the CG algorithm is not applicable. In such cases, the Generalized Minimum Residual (GMRES) solver is popular. It is also a Krylov Subspace solver that, as the name implies, tries to minimize the 2-norm of the residual of the linear system in each iteration \cite{gmres_1986}.
However, unlike in the symmetric case, there is no `short-term' recurrence formula, which means previous iterations' Krylov basis vectors need to be stored. This significantly increases the memory requirement.
Certain variants, such as using CGS2 reorthogonalization (see section \ref{sec:hpgmp}), also use some dense BLAS2 routines.

Furthermore, HPCG is required to use double-precision arithmetic. As stated in the introduction, it is of interest to find mixed-precision algorithms for PDE-based and other simulation workloads, and it is thus of interest to measure HPC systems' expected performance on such workloads.
Research in mixed-precision methods for sparse matrices is not new. In 2008, Buttari et al. \cite{buttari_sparse_2008} investigated mixed-precision CG and GMRES solvers, among other mixed-precision solvers for sparse problems. More recently, Loe et al. \cite{loe_mxp_gmres_2021} implemented mixed-precision GMRES using two different strategies - starting a single-precision solver and then switching to double after some iterations, and iterative refinement (GMRES-IR), described in the next section.
They used either polynomial preconditioning or block-Jacobi preconditioning, whose characteristics in terms of preconditioning effectiveness, parallelism and resource utilization are quite different from the multigrid preconditioner prescribed by HPG-MxP.
They test their implementation using sample sparse matrices from applications on a single NVIDIA V100 GPU.

HPG-MxP, then known as the High Performance GMRES Multiprecision (HPGMP) benchmark, was first proposed in 2022 \cite{hpgmp}. It solves a similar problem as the HPCG benchmark, but as the name suggests, uses a GMRES solver and allows the use of lower precisions.

As far as the authors are aware, Anzt et al. \cite{xsdk_mxp_kit} were the first to run HPG-MxP on the Frontier exascale system at Oak Ridge National Laboratory.

\section{The HPG-MxP benchmark}
\label{sec:hpgmp}

HPG-MxP, introduced as HPGMP by Yamazaki et al. \cite{hpgmp} in 2022, is the first benchmark to target mixed-precision simulation workloads.
Similar to HPCG, it solves a Poisson-like problem on a uniform mesh. This time, there is an option to make the problem nonsymmetric, though as Yamazaki et al. observe, for GMRES the symmetric version is at least as difficult to solve as the nonsymmetric one.
The matrix they construct is weakly diagonally dominant; that is, for each row $i$, $\sum_{j \neq i} |a_{ij}| \leq a_{ii}$.

The original right-preconditioned GMRES algorithm \cite[chapter~9]{saadbook} is shown in algorithm \ref{alg:gmres_orig}.
It attempts to minimize the 2-norm of the residual, and generation of orthogonal Krylov basis vectors is necessary for this. The Arnoldi process used to build the orthogonal Kyrlov basis may use one of a few different methods, such as classical Gram-Schmidt and modified Gram-Schmidt.
While classical Gram Schmidt is amenable to more efficient implementation, it is more prone to round-off errors and resulting loss of orthogonality.

\begin{algorithm}
    \begin{algorithmic}
        \Require Initial guess $\bld{x}_0$
        \State $\bld{M} \gets \mathcal{P}(\bld{A})$
        \Comment Preconditioner generation
        \State $\bld{r}_0 \gets \bld{b}-\bld{Ax}_0$, $\beta = ||\bld{r}_0||$, $\bld{q}_1 = \bld{r}_0/\beta$.
        \For {$j = 1,2,...,m$ do}
            \State $\bld{w} \gets \bld{AM}^{-1}\bld{q}_j$
            \Comment Preconditioner application and SpMV
            \For{$i = 1,2,3,...,j$}
                \State $h_{i,j} \gets (\bld{w}, \bld{q}_i)$
                \State $\bld{w} \gets \bld{w} - h_{i,j}\bld{q}_i$
            \EndFor
            \State $h_{j+1,j} \gets \lVert\bld{w} \rVert_2$, $\bld{v}_{j+1} \gets \bld{w} / h_{j+1,j}$
            \State $V_m := [\bld{v}_{1},...,\bld{v}_m]$, $\bar{\bld{H}}_m \gets \{h_{i,j}\}_{1\leq i\leq j; 1\leq j\leq m}$
        \EndFor
        \State $\bld{y}_{m} \gets \mathrm{argmin}_y\lVert \beta \bld{e}_1- \bar{\bld{H}}_m\bld{y}\rVert_2$
        \State $\bld{x}_m \gets \bld{x}_0 + \bld{M}^{-1}\bld{Q}_m\bld{y}_m$
        \If{$||\bld{b}-\bld{Ax}_m||_2 < \tau$}
            \State Exit.
        \Else
            \State Restart with $\bld{x}_0 \gets \bld{x}_m$.
        \EndIf
    \end{algorithmic}
    \caption{Right-preconditioned GMRES algorithm for a nonsingular linear system $\bld{Ax}=\bld{b}$}
    \label{alg:gmres_orig}
\end{algorithm}

Algorithm \ref{alg:gmres_orig} is called `right-preconditioned' since the preconditioner applies on the right - it is an algorithm to solve the linear system
$\bld{AM}^{-1}y = \bld{b}$, whose solution $\bld{x} = \bld{M}^{-1}\bld{y}$ is the same as that for $\bld{Ax} = \bld{b}$.
The preconditioner is one cycle of geometric multigrid with a forward Gauss-Seidel smoother. The restriction $\bld{R}$ is a simple injection from every alternative fine grid point, while the prolongation operator is its transpose $\bld{P} = \bld{R}^T$. If $c_f(i)$ is the index of the $i$th coarse grid point in the fine grid and $\bld{v}$ is a fine grid vector,
\begin{equation}
    (\bld{Rv})_i := v_{c_f(i)}.
    \label{eqn:restrict}
\end{equation}

In the HPG-MxP benchmark, mixed-precision is utilized at the highest level via the idea of iterative refinement applied to GMRES, GMRES-IR.
Since GMRES attempts to generate orthogonal Krylov basis vectors, loss of orthogonality is detrimental to its convergence \cite{giraud_orthogonality_2005}.
This is especially a problem with classical Gram-Schmidt orthogonalization and the usage of lower precisions.
Therefore, the version of GMRES-IR prescribed by the benchmark uses reorthogonalization steps to better preserve orthogonality.

\begin{algorithm}
    \begin{algorithmic}[1]
        \Require Initial guess $\bld{x}_0$, restart length $m$, tolerance $\tau$.
        \State $\bld{M} \gets \mathcal{P}_{mg}(\bld{A})$
        \Comment Multigrid preconditioner generation
        \State $\rho_0 \gets \lVert \bld{b} \rVert_2$
        \State {\color{blue} $\bld{t} \in \mathbb{R}^{m+1}, \bld{c} \in \mathbb{R}^{m+1}, \bld{s} \in \mathbb{R}^{m+1}$}
        \State {\color{blue} $\bld{H} \in \mathbb{R}^{(m+1)\times m}$}
        \Comment Projected system matrix
        \State {\color{blue} $\bld{Q} = [\bld{q}_1, \bld{q}_2, ..., \bld{q}_m] \in \mathbb{R}^{n\times m}$}
        \Comment Krylov basis vectors
        \While{not converged}
            \State $\bld{r} \gets \bld{b}-\bld{Ax}$, $\rho = ||\bld{r}||_2$
            \If{$\rho / \rho_0 < \tau$}
                \State Converged, break.
            \EndIf
            \State $\bld{r} \gets \bld{r} / \rho$
            \State {\color{blue}$\bld{q}_1 \gets $} $\bld{r}$
            %\State {\color{blue} $h_{1,1} \gets 0$
            \State {\color{blue} $t_{0,0} \gets$} $\rho$.
            \For {$k = 1,2,...,m$ do}
                \If{$\rho/\rho_0 < \tau$}
                    \State Converged, break.
                \EndIf
                \color{blue}
                \State $\bld{z} \gets \bld{M}^{-1}\bld{q}_k$
                \Comment Multigrid preconditioner
                \State $\bld{q}_{k+1} \gets \bld{A}\bld{z}$
                \Comment SpMV to get next basis vector
                \Procedure{CGS2 orthogonalization}{$\bld{q}_{k+1}$}
                    \State $\bld{h} \gets \bld{Q}_{[1:k]}^T \bld{q}_{k+1} \in \mathbb{R}^k$
                    \Comment GEMVT
                    \State $\bld{q}_{k+1} \gets \bld{q}_{k+1} - \bld{Q}_{[1:k]} \bld{h}$
                    \Comment GEMV
                    \State $H_{1:k,k} \gets \bld{h}$
                    \State $\bld{h} \gets \bld{Q}_{[1:k]}^T \bld{q}_{k+1}$
                    \Comment reorth. GEMVT
                    \State $\bld{q}_{k+1} \gets \bld{q}_{k+1} - \bld{Q}_{[1:k]} \bld{h}$
                    \Comment reorth. GEMV
                    \State $H_{1:k,k} \gets H_{1:k,k} + \bld{h}$
                \EndProcedure
                \State $\beta \gets \lVert \bld{q}_{k+1} \rVert_2$
                \State $\bld{q}_{k+1} \gets \bld{q}_{k+} / \beta $
                \Comment Normalize the new basis vector
                \State $H_{k+1,k} \gets \beta$
                \Procedure{Update QR with Given's rotations}{}
                    \For{$j = 1,2,...,k-1$}
                        \State $H_{j+1,k} \gets -s_j H_{j,k} + c_j H_{j+1,k}$
                        \State $H_{j,k} \gets c_j H_{j,k} + s_j H_{j+1,k}$
                    \EndFor
                    \State $\mu \gets \sqrt{H_{k,k}^2 + H_{k+1,k}^2}$
                    \State $H_{k,k} \gets \mu$
                    \State $H_{k+1,k} \gets 0$
                    
                    \State $\rho \gets |t_k s_j|$.
                    
                    \State $t_{k+1} \gets -t_k s_j$
                    \State $t_k \gets t_k c_j$
                    \State $s_k \gets s_j$, $c_k \gets c_j$
                \EndProcedure
            \color{black}
            \EndFor % k loop
            \Comment Restart cycle completed
            \color{blue}
            \State $\bld{t} \gets \bld{H}^{-1}\bld{t}$
            \Comment Dense TRSM of size $m$
            \State $\bld{r} \gets \bld{Qt}$
            \color{black}
            \State $\bld{x}_m \gets \bld{x}_0 + {\color{blue} \bld{M}^{-1}\bld{r}}$
            \Comment Mixed-precision update
        \EndWhile
    \end{algorithmic}
    \caption{Right-preconditioned mixed-precision GMRES-IR for the linear system $\bld{Ax}=\bld{b}$. Steps in blue are allowed to be performed in low or mixed precision.}
    \label{alg:gmres_ir_cgs2}
\end{algorithm}

We show the GMRES-IR CGS2 algorithm \ref{alg:gmres_ir_cgs2} used by the benchmark, including details of how the least-squares problem (QR factorization) is solved using Given's rotations.
The key aspect to note here is that many of the operations are allowed to be performed in low precision, but the residual update in line 7 and solution update in line 47 are required to be done in double-precision. This makes it possible for the solution to converge to an accuracy level equivalent to that of a fully double precision solver.
Note that the QR factorization update using Given's rotations happens on the CPU, on each process redundantly.

Though they provide the option of using a nonsymmetric matrix for the benchmark, Yamazaki et al. \cite{hpgmp} prefer the same symmetric weakly diagonal-dominant matrix as in HPCG, since this matrix actually takes more iterations for GMRES to converge than their nonsymmetric variant.
The matrix has all diagonal entries equal to 26 and all off-diagonal entries equal to -1.

The benchmark consists of three phases:
\begin{enumerate}
    \item validation,
    \item mixed-precision benchmark, and
    \item double-precision `reference' benchmark.
\end{enumerate}

The benchmark first validates the mixed-precision solver on a few, fixed number of processors. By default, this is all the GPUs on one node.
The validation consists of first running the double precision GMRES solver, starting from a zero initial guess, to converge the residual norm by 9 orders of magnitude. The number of iterations required, $n_d$, is recorded.
Next, mixed-precision GMRES-IR is run to converge to the same tolerance, starting from a zero initial guess again. The number of iterations required, $n_{ir}$, is recorded.

This is followed by the benchmark phase, where mixed-precision GMRES-IR is first run for a fixed number of iterations. The time taken by each motif is recorded and the number of floating point operations is counted using a carefully constructed model. Floating point operations of different precisions are counted equally, and thus the reported GFLOP/s number is a mixed-precision number, not the standard double-precision GFLOP/s.
The final GFLOP/s metric is penalized by the ratio of the iteration counts obtained in the validation phase.
That is, the final floating point throughput $F$ is given by $F = F_{\mathrm{raw}} \frac{n_d}{n_{ir}}$.
Thus, to avoid a heavy penalty, mixed-precision GMRES-IR must not take too many additional iterations to converge 9 orders of magnitude. It is in this sense that the mixed-precision solver is required to provide a solution that is `somewhat close' to that of the double-precision solver.
When the ratio $\frac{n_d}{n_{ir}}$ is less than 1, it is multiplied by the final GFLOPS value to penalize the mixed precision run.
However, then the ratio is greater than 1, no penalty is applied.
Thus, in the event that the mixed precision solver actually takes \emph{fewer} iterations to converge for any reason, this is not regarded as an advantage for the mixed precision solver, and it is regarded as though the mixed precision solver has the same convergence rate as double precision GMRES.

The GMRES-IR solution process is repeated, starting from a zero initial guess each time, until the requested running time is filled.
Similar to HPCG, the official running time proposed by Yamazaki et al. \cite{hpgmp} is 1800 seconds.
Results from this phase are labelled as `mxp' in the results section.

Finally, a double-precision GMRES solver is run for the same number of fixed maximum iterations, and similar performance metrics are collected. We report results from this phase labelled as the `double' in the results section.

\subsection{The reference implementation}
The reference implementation used in this work is from the official HPG-MxP repository \footnote{https://github.com/hpg-mxp/hpg-mxp} as of April 2025.
As described by Yamazaki et al. \cite{hpgmp}, there are many substantial inefficiencies in this version. This is understandable since their aim was to propose the benchmark, not provide a fully optimized implementation.
However, we aim for an optimized implementation, for which we document the issues with the reference implementation first.
\begin{enumerate}
\item The Gauss-Seidel implementation uses SpTRSV from cuSparse and rocsparse, which use a level-scheduled implementation \cite{naumov_sptrsv_2011} without reordering. This variant is mathematically equivalent to a sequential lexicographic (ordered spatially by mesh points) Gauss-Seidel, but it does not expose the most parallelism and does not fully utilize the GPU \cite{suchoski_parallel_triangular_2012}.
\item Forward Gauss-Seidel is performed using a separate SpMV with the $\bld{U}$ matrix followed by a SpTRSV with the $\bld{L}$ matrix, which is wasteful.
\item In multigrid, following the Gauss-Seidel smoother, the smoothed fine-grid residual is first computed using SpMV, followed by restriction of this to the coarse grid. This performed substantial additional work which is not necessary, since the restriction is simple injection of values from selected fine-grid points to the coarse grid.
\item There is no overlap of communication and computation in either the SpMV or the Gauss-Seidel operation. Indeed, the code does not support any asynchronous behaviour.
\item The sparse matrix format used is compressed sparse row, or CSR. While this format is a popular choice and reasonably efficient in general, on GPU architectures it has been shown that for stencil-based problems like the one here, other formats can work better \cite{bell_spmv_2009}.
\item All mixed-precision operations are done on the host, necessitating additional host-device copies and utilizing slower CPU DRAM.
\end{enumerate}

\subsection{Optimizations}
\label{sec:opt}

In response to the issues identified with the reference implementation, we carry out several algorithmic changes and optimizations.

\subsubsection{Multicolor Gauss-Seidel iteration}
To start with, we implement the forward Gauss-Seidel in its `relaxation' version \cite{async_thesis_2020}, completing the operation in one sweep over the matrix.

More importantly, we reorder the matrix and vectors symmetrically using an independent set ordering in order to expose fine-grained parallel work in the Gauss-Seidel kernel on GPUs.
Each subdomain is reordered independently, without any communication.
If the $n$ local mesh points can be divided into $n_c$ independent sets such that no two points within a set are directly connected to each other in the sparsity pattern of the system matrix, a Gauss-Seidel iteration can be completed in $n_c$ operations one after the other, each working on $n/n_c$ rows in a fully parallel manner.
If $n_c = \mathcal{O}(1)$ independent of $n$, we should be able to achieve good parallel efficiency on a GPU.

The ordering itself is computed on the GPUs using the Jones-Plassmann-Luby (JPL) algorithm \cite{luby_coloring_1986,jones_plassmann_coloring_1993} in the optimization function of the benchmark.
Naumov et al. introduced a GPU implementation \cite{naumov_coloring_ilu_2015}. We use the implementation by Trost et al. \cite{rochpcg}.

\def\hh{0.5}% step size for grid
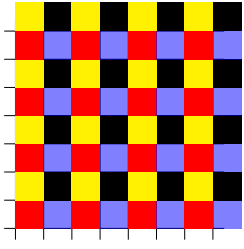
\begin{figure}
\begin{tikzpicture}[scale=0.75]
    \draw[step=\hh,very thin] (-2.2,-2.2) grid (1.7,1.7);
	%\node at (-0.25,-0.25) {\tiny $ij$};
    \foreach \i in {0,...,3}{
        \foreach \j in {0,...,3}{
        	\fill [red] (-2+\i,-2+\j) rectangle (-1.5+\i,-1.5+\j);
        	\fill [blue,semitransparent] (-1.5+\i,-2+\j) rectangle (-1.0+\i,-1.5+\j);
        	\fill [black] (-1.5+\i,-1.5+\j) rectangle (-1.0+\i,-1.0+\j);
        	\fill [yellow] (-2+\i,-1.5+\j) rectangle (-1.5+\i,-1.0+\j);
         }
     }
	%\fill [blue,semitransparent] (-0.5,-0.5) rectangle (0,0);
\end{tikzpicture}
\caption{An example of independent sets for a 2D 9-point stencil, the 2D analog of the 3D 27-point stencil used in HPCG and HPG-MxP}
\label{fig:multicolor}
\end{figure}

As shown in figure \ref{fig:multicolor}, the JPL algorithm as well as a sequential greedy algorithm \cite[section~3.3.3]{saadbook} applied to a 9-point stencil on a spatially two-dimensional (2D) mesh give 4 independent sets, or `colors'.
The analogous 27-point stencil in 3D requires 8 colors.
This enables the use of multicolored Gauss-Seidel, which involves fully parallel operations on the rows within each color.
The convergence rate sometimes suffers compared to lexicographic Gauss-Seidel and other types of reorderings such as Reverse Cuthill McKee (RCM) \cite{benzi_orderings_1999,async_thesis_2020}.
However, this is less of an issue within a multigrid preconditioner setting.

\subsubsection{ELLPACK matrix format}

\begin{figure}
	\begin{subfigure}{0.49\linewidth}
		\centering
    \resizebox{0.7\columnwidth}{!}{%
		\begin{tikzpicture}
		\def \nx{5}
		\def \ny{8} %
		\def \ymax{6} %
		\draw (0.5,\ny+0.5) node[right] {Stencil entries $\rightarrow$};
		\draw (-0.5,\ny/2) node[rotate=270] {Rows (mesh points) $\rightarrow$};
		\draw (0,0) grid (\nx-1,2);
		\draw (0,2) grid (\nx,\ny-1);
		\draw (0,\ny-1) grid (\nx-2,\ny);
		
		\draw[->,thick,blue] (0.5,\ny-1+0.5) -- (\nx-2.5,\ny-1+0.5);
		\draw[->,thick,dashed,blue] (\nx-2.5,\ny-1+0.5) -- (0.5,\ny-1-0.5);
		
		\foreach  \l in {2,...,\ymax}
		{
			\draw[->,thick,blue] (0.5,\ny-\l+0.5) -- (\nx-0.5,\ny-\l+0.5);
			\ifnum\l<\ny
			\draw[->,thick,dashed,blue] (\nx-0.5,\ny-\l+0.5) -- (0.5,\ny-\l-0.5);
			\fi
		}
		\foreach  \l in {7,...,\ny}
		{
			\draw[->,thick,blue] (0.5,\ny-\l+0.5) -- (\nx-1.5,\ny-\l+0.5);
			\ifnum\l<\ny
			\draw[->,thick,dashed,blue] (\nx-1.5,\ny-\l+0.5) -- (0.5,\ny-\l-0.5);
			\fi
		}
		\end{tikzpicture}
      }
		\caption{CSR}
	\end{subfigure}
	\begin{subfigure}{0.49\linewidth}
		\centering
    \resizebox{0.7\columnwidth}{!}{%
		\begin{tikzpicture}
		\def \nx{5}
		\def \ny{8} %
		\draw (0.5,\ny+0.5) node[right] {Stencil entries $\rightarrow$};
		\draw (-0.5,\ny/2) node[rotate=270] {Rows (cells) $\rightarrow$};
		\draw (0,0) grid (\nx,\ny);
		\foreach  \l in {1,...,\nx}
		{
			\draw[->,thick,blue] (\l-0.5,\ny-0.5) -- (\l-0.5,0.5);
			\ifnum\l<\nx
			\draw[->,thick,dashed,blue] (\l-0.5,0.5) -- (\l+0.5,\ny-0.5);
			\fi
		}
		\end{tikzpicture}
      }
		\caption{ELL}
	\end{subfigure}
	\caption{Two possible layouts of the non-zero coefficients' array of a matrix}
	\label{fig:layouts}
    \vspace{-0.75cm}
\end{figure}
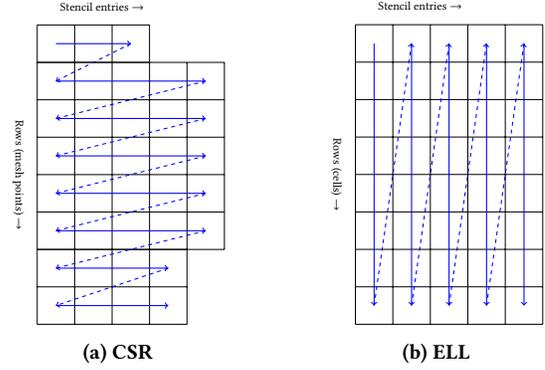

We use the ELLPACK matrix format \cite{bell_spmv_2009}, sometimes simply referred to as ELL. This format is able to fully utilize GPU warps, especially the 64-wide warps of AMD GPUs, when there are only a few non-zeros per row. Fully utilizing the warps is more difficult using the CSR format \cite[section~2.9]{async_thesis_2020}, which is used in the reference implementation of HPG-MxP by Yamazaki et al. \cite{hpgmp}.
The memory layouts used by the two formats are depicted in figure \ref{fig:layouts}.
While the ELL format may have some overhead in its values and column-indices array due to padding in rows having less than the maximum number of nonzeros, it does not require a row pointer array.

\subsubsection{Compute-communication overlap}
Similar to rocHPCG \cite{rochpcg}, we implement SpMV and Gauss-Seidel operations that update the interior points while neighborhood halo communication operations take place asynchronously.
Pre-conditions for each operation are ensured using two GPU streams (`compute' stream and `halo' stream) and one event per MPI rank.
The non-blocking kernels that compute the matrix-vector product for the both the interior+boundary and halo regions are enqueued on the `compute' stream, while the non-blocking buffer packing kernel and asynchronous host-device copies (if used) are enqueued on the `halo' stream.

The event is the object that achieves precise synchronization between the two streams during the Gauss-Seidel operation. This operation has a significant difference from the SpMV operation - unlike in SpMV, Gauss-Seidel (equation \eqref{eqn:fgs}) requires the output vector $\bld{y}$ to be communicated, not the input vector.
The event is used to ensure that the interior computation kernel begins only after boundary entries of the initial vector $\bld{y}$ have been copied into the send buffer, since the interior kernel updates boundary entries in addition to fully interior locations.

\subsubsection{Fused SpMV-restriction}
The original implementation of Yamazaki et al. \cite{hpgmp} explicitly stores the restriction matrix $\bld{R}$. Representing the fine grid index set as $F$ and the coarse grid by $C$, the coarse grid residual is computed as
\begin{align}
    v_i &\gets \sum_{j \in \mathcal{N}_i} A_{ij}^f x_j^f, \quad \forall\, i \in F \\
    r_i^c &\gets \sum_{j \in \mathcal{N}_i} R_{ij} (b_j - v_j) \quad \forall\, i \in C.
\end{align}
where $\bld{b}$ is the right-hand side vector and $\bld{x}^f$ is the pre-smoothed solution vector on the fine grid.
Since the restriction is a simple injection as shown in equation \eqref{eqn:restrict}, we can fuse the residual calculation and restriction operations.
Thus, in our implementation, we compute the smoothed residual only at the coarse grid points, rather than separately compute it at all fine grid points and then restrict to the coarse grid. Using the coarse-to-fine grid index mapping $f_c$,
\begin{equation}
    r_i^c \gets b_{f_c(i)} - \sum_{j \in \mathcal{N}(f_c(i))} A_{f_c(i)j}^f x_j^f \quad \forall\, i \in C.
\end{equation}
Note that we do not store the restriction operator explicitly.
We updated the accounting of the number of floating point operations in the multigrid preconditioner to include this optimization.

\subsubsection{Software engineering}
\label{sec:software}
Given that the reference implementation is cross-platform and runs on both AMD and NVIDIA GPUs, we preserved this aspect and made sure that our implementation works with high performance on both vendors' devices.
Taking inspiration from numerical libraries like Ginkgo \cite{ginkgo_2022}, we introduce a device context \texttt{DeviceCtx} that abstracts many vendor-specific details, including device memory allocation and deallocation, GPU stream and event operations, initialization and destruction of BLAS and sparse BLAS library handles etc.
We use C++ features like function overloading and templates to generate high-performance kernels for both CUDA and AMD devices. For example, in sparse matrix-vector product, values loaded from the input vector are constant and may be used more than once, and it makes sense to cache them in L1 cache.
However, the matrix nonzeros are read only once, it make sense to skip temporal caching for these values and improve performance slightly.
The intrinsic (backend-specific, non-standard) functions for these types of loads and stores are different for the two platforms and we take this into account using C++ features.

Note that all of the improvements detailed so far apply to both the mixed-precision and purely double-precision solvers.

Additionally, we implement simple custom mixed-precision GPU kernels for operations such as WAXPBY as required by the mixed-precision GMRES-IR implementation. This allows us to remove the host-device copies performed by the reference implementation to perform these operations on the CPU.

\subsection{Alternative full-scale validation}
\label{sec:alt_valid}

Yamazaki et al. argue \cite{hpgmp} that validating the mixed-precision GMRES-IR on a small number of processes (typically 1 node) and corresponding small problem size is sufficient.
The reason given is that the benchmark's purpose is to measure the computer's ability to perform operations representative of typical HPC applications while allowing the use of different precision formats, not to provide a truly scalable solver.
However, as they themselves note, when the multigrid preconditioner is used, the loss of convergence rate when using mixed precision GMRES-IR can be worse at larger scales.
In order to investigate the impact of validation at a 1-node scale on the penalty factor, we introduce a new validation mode in our version of the benchmark code.

In the new validation mode, all the processes available to the run and used for the benchmarking phase, are also used for the validation phase. The global problem size used for the two phases is also the same. Two modes are provided:
\begin{enumerate}
\item \texttt{standard}: Double-precision GMRES is run on a small subset of processes, 1 node, until a relative residual norm of $10^{-9}$ is reached. Since the problem size is correspondingly small, this always happens within the iteration limit of 10,000.
Mixed-precision GMRES-IR is then also converged 9 orders of magnitude and the number of iterations $n_{ir}$ is recorded.
\item \texttt{fullscale}: Double precision GMRES is run for a maximum of $n_d$ (10,000) iterations or a relative residual norm of $10^{-9}$, whichever comes first. The achieved relative residual norm $\tau$ is recorded. Mixed precision GMRES-IR is then run and converged until the same relative residual norm $\tau$ is achieved, and the number of iterations $n_{ir}$ is recorded.
\end{enumerate}
As Yamazaki et al. \cite{hpgmp} noted, GMRES takes more and more iterations to converge to a fixed tolerance as the problem scale increases.
Thus, with our new validation path, at low scales, the GMRES solver hits the $10^{-9}$ tolerance much before it reaches 10,000 iterations.
The mixed precision GMRES-IR is then required to converge to $10^{-9}$.
However, at large scales, the global problem size is much larger and GMRES hits 10,000 iterations first. Eg., at 1024 nodes in our runs, the solver achieves a relative residual of about $1.15\times 10^{-5}$ at 1024 nodes.
This was chosen in order to cap the amount of time the whole benchmark takes at very large scales, while still learning something about any loss of convergence caused by the use of mixed precision.

%\begin{table*}
%    \begin{tabular}{|c|c|c|}
%    \hline
%    Aspect  &  Yamazaki et al. & Present \\
%    \hline
%    Matrix format & compressed sparse row & ELLPACK \\
%    Gauss-Seidel smoother & Lexicographic, level-scheduled cuSparse/rocSparse & Multicolored \\
%    Communication & Halo exchange followed by all computations & Halo exchange overlapped with interior computations \\
%    \hline
%    \end{tabular}
%    \caption{Differences between the present implementation and the reference implementation of Yamazaki et al. \cite{hpgmp}}
%\end{table*}

\section{Results}

We ran the optimized code on the Frontier system at Oak Ridge National Laboratory using AMD ROCm 6.2.4, Cray MPICH 8.1.31 and GCC 14.2.
Each node consists of a 64-core AMD Milan CPU and 4 AMD MI250x GPUs, each divided into two Graphics Compute Dies (GCDs). Each GCD is effectively treated as a separate GPU. Thus, we consider there to be 8 GPUs per node.
Each GCD is equipped with 64 GB of High Bandwidth Memory (HBM) with a vendor-claimed peak bandwidth of 1.6 TB/s. This HBM is generally referred to as the `global' memory of the GPU device, as opposed to its much smaller but faster L2 and L1 caches.
The CPU portions of the code (problem generation etc.) utilize OpenMP parallelism as in the reference code.

Table \ref{tab:parameters} summarizes the parameters we used to run the benchmark. We use a restart length of 30, similar to Yamazaki et al. \cite{hpgmp}. This is also the default restart length in the popular PETSc package \cite{petsc-user-ref}.
\begin{table}
    \centering
    \begin{tabular}{|c|c|}
    \hline
    Parameter & Value \\
    \hline
    Restart length & 30 \\
    Local mesh size & $320^3$ \\
    Specified running time (< 1024 nodes) & 1800 s \\
    Specified running time (>= 1024 nodes) & 900 s \\
    Max. GMRES iterations per solve & 300 \\
    No. GCDs used for validation & 8 \\
    Relative convergence tolerance for validation & 1e-9 \\
    \hline
    \end{tabular}
    \caption{HPG-MxP parameters used}
    \label{tab:parameters}
    \vspace{-1cm}
\end{table}

Anzt et al. \cite{xsdk_mxp_kit} ran the reference version of the code on Frontier from 1 to 8192 nodes, using the reference implementation of Yamazaki et al. Due to the inefficiencies in this implementation detailed earlier, we do not expect it to give the best performance.
However, as of writing, it is the state of the art in HPG-MxP performance on Frontier, so we include it in our results.
Please note that in the graphs that follow, the points corresponding to the results by Anzt et al. \cite{xsdk_mxp_kit}, labelled ``xsdk", are somewhat approximate as they have been read off a graph (figure 4 under the chapter `Advances in mixed precision algorithms').

During validation on 1 node (8 GCDs), the reference double-precision GMRES solver takes 2305 iterations to converge 9 orders of magnitude, while the mixed-precision GMRES-IR requires 2382 iterations to converge to the same tolerance.
This small increase in the required number of iterations is expected, and the appropriate penalty is applied to the mixed-precision performance metric.

\subsection{Scaling, speedup and full-system performance}

\begin{figure}
    \includegraphics[width=0.98\linewidth]{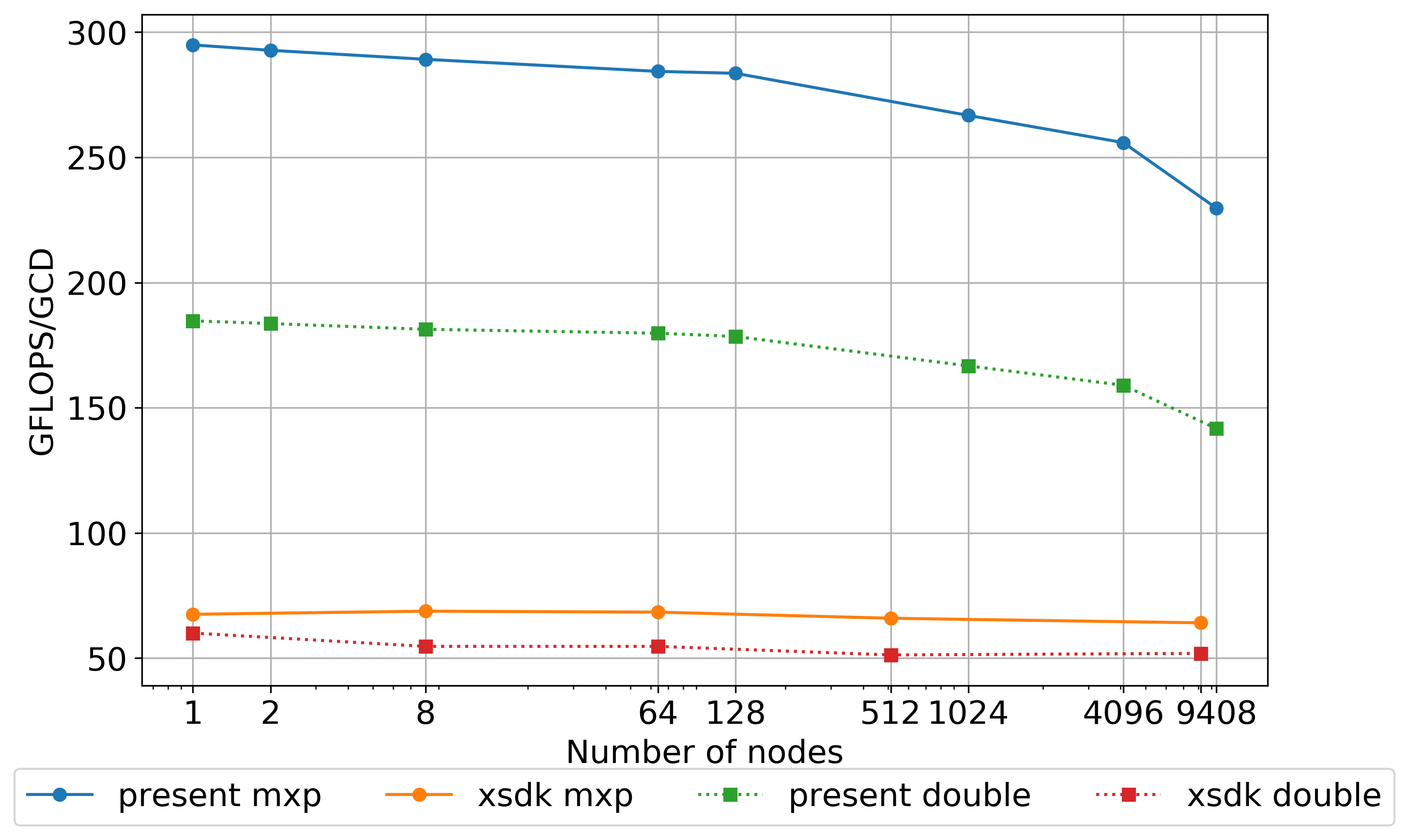}
    \caption{`Weak' scaling of the overall benchmark on Frontier. It is based on the penalized time taken by the mixed-precision solver to complete the specified number of iterations.}
    \label{fig:weak_scaling}
\end{figure}

We first discuss the scaling results of the benchmarking phase. Figure \ref{fig:weak_scaling} shows how the overall performance per GCD scales as we increase the problem size and the number of GCDs in proportion, for both our implementation (`present') and the reference implementation (`xsdk').
This is similar to weak scaling, though we do not regard this as true weak scaling because the solver does not converge to a specified residual tolerance, but rather executes a fixed number of iterations.
We see that the performance holds up well up to large scales. However, as we approach the full system scale, the scaling efficiency decreases due to the many inner products required by the GMRES algorithm. Each inner product requires a global all-reduce operation.
Even though the CGS2 version batches the inner product into a transposed GEMV operation and thus reduces the effective latency, we still see some degradation of the overall scaling.
Depending on the mapping of subdomains to MPI ranks, the coarse multigrid levels may also contribute to some of this decrease in efficiency (see the discussion on tracing below and figure \ref{fig:trace}).
Since the reference implementation achieves much lower performance in general, it does not see this effect.
The weak scaling efficiency of our implementation from 1 node to 9408 nodes is 78\%.

At the full system scale of 9408 nodes or 75,264 GPUs, we achieve an overall mixed-precision performance of 17.23 petaflops.
For perspective,
%the number reported for Frontier on the HPCG benchmark in the TOP500 list is 14.05 petaflops, though these are not directly comparable.
when we ran HPCG ourselves on Frontier on 9408 nodes, we achieved 10.4 petaflops.
We note that these numbers are not directly comparable since the solvers are different in the two benchmarks.

\begin{figure}
    \begin{subfigure}{0.99\linewidth}
        \includegraphics[width=0.98\linewidth]{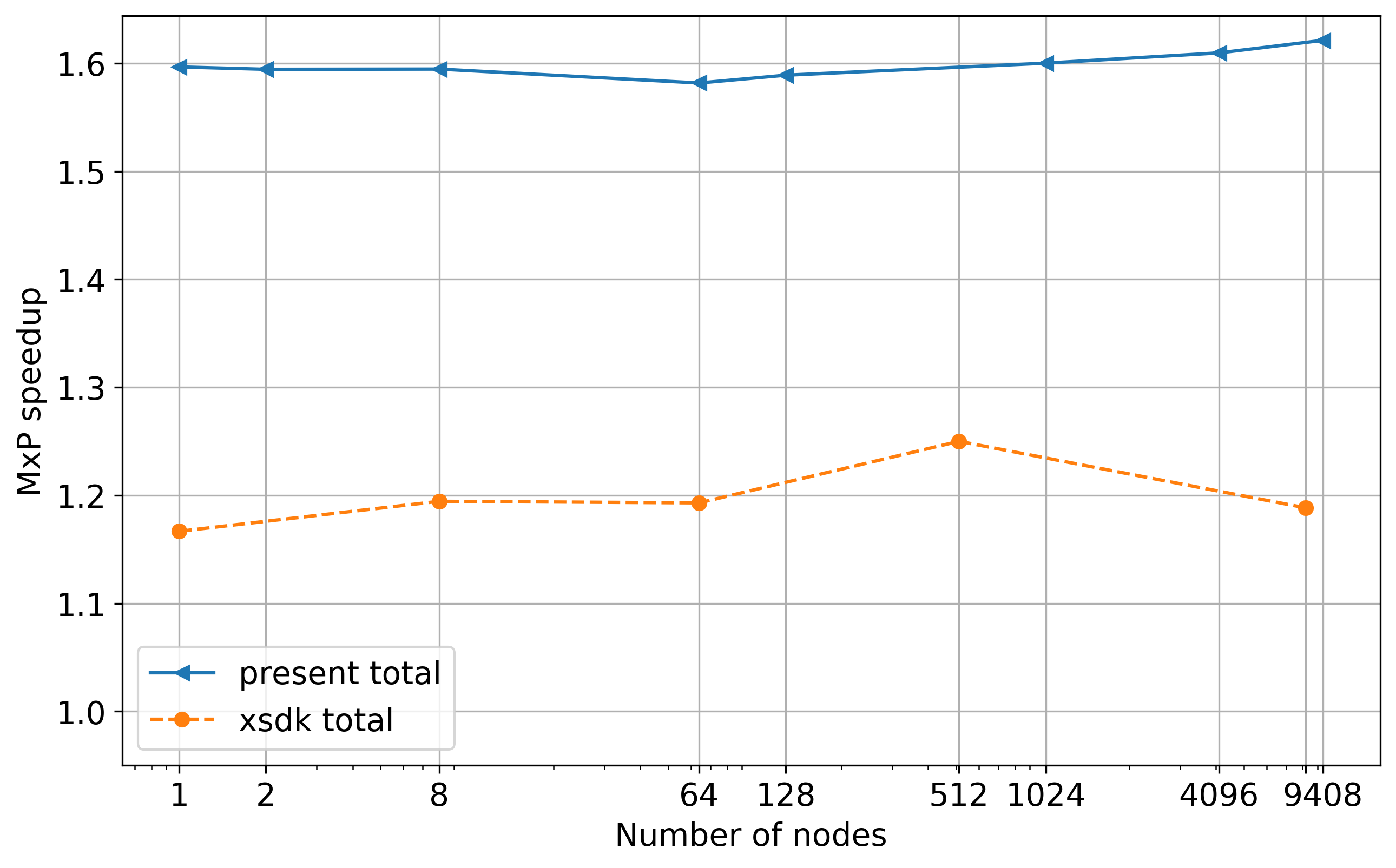}
    \end{subfigure}
    \begin{subfigure}{0.99\linewidth}
        \includegraphics[width=0.98\linewidth]{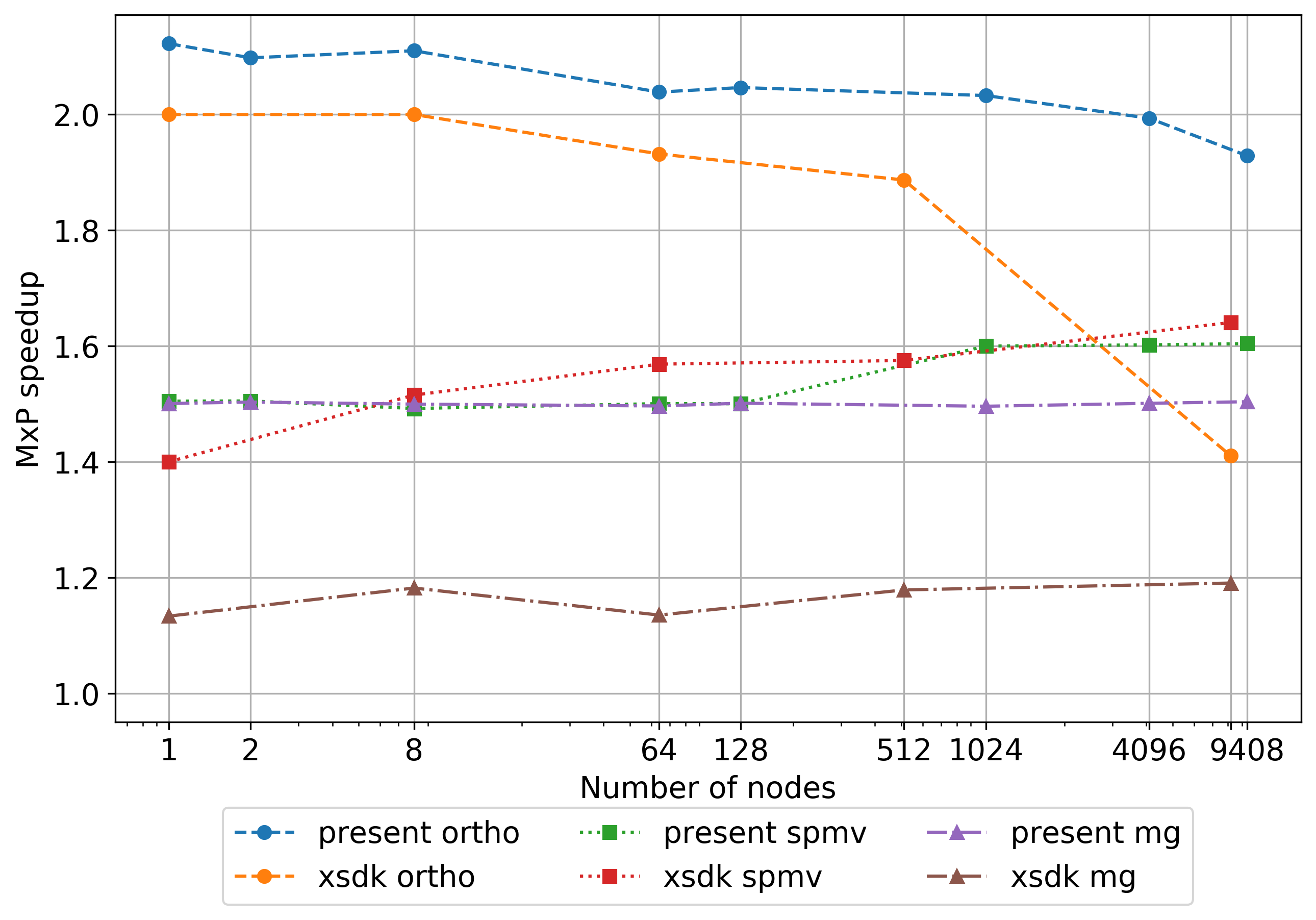}
    \end{subfigure}
    \caption{Speedups obtained on Frontier for different computational motifs in mixed-precision GMRES-IR over double precision GMRES. They are based on the penalized GFLOP/s rating of the mixed-precision solver to complete the specified number of iterations compared to that of the double-precision solver.}
\label{fig:speedup}
\end{figure}

Figure \ref{fig:speedup} shows the (penalized) speedup obtained by mixed single-double precision GMRES-IR versus double precision GMRES.
We see a remarkable overall speedup of about 1.6$\times$, against a theoretical peak of 2$\times$ for going fully to single precision assuming the code is limited by memory bandwidth.
This is much improved compared to the speedups obtained using the reference implementation.
Interestingly, about 1.6$\times$ was also the speedup reported by Buttari et al. \cite{buttari_sparse_2008} for mixed double-single precision GMRES on a single Intel Woodcrest CPU from the year 2006.

Clearly, the perfect speedup of the orthogonalization phase plays a role in this. Since this operation is a dense BLAS-2 operation, it makes the best use of increased memory throughput of lower precision numbers.
At very large scales, however, the orthogonalization spends more time in MPI all-reduce operations, thus reducing the speedup somewhat.
Multigrid (primarily Gauss-Seidel, as we shall see) and SpMV drag the speedup down somewhat owing to their the need to fetch index arrays, leading to lower arithmetic intensity and lower advantage from decreasing the bit-width of the floating point numbers.
We note that optimizing the motifs in multigrid and SpMV as detailed in subsection \ref{sec:opt} significantly improves the attained speedup.

\begin{figure}
    \begin{subfigure}{0.99\linewidth}
        \includegraphics[width=0.98\linewidth]{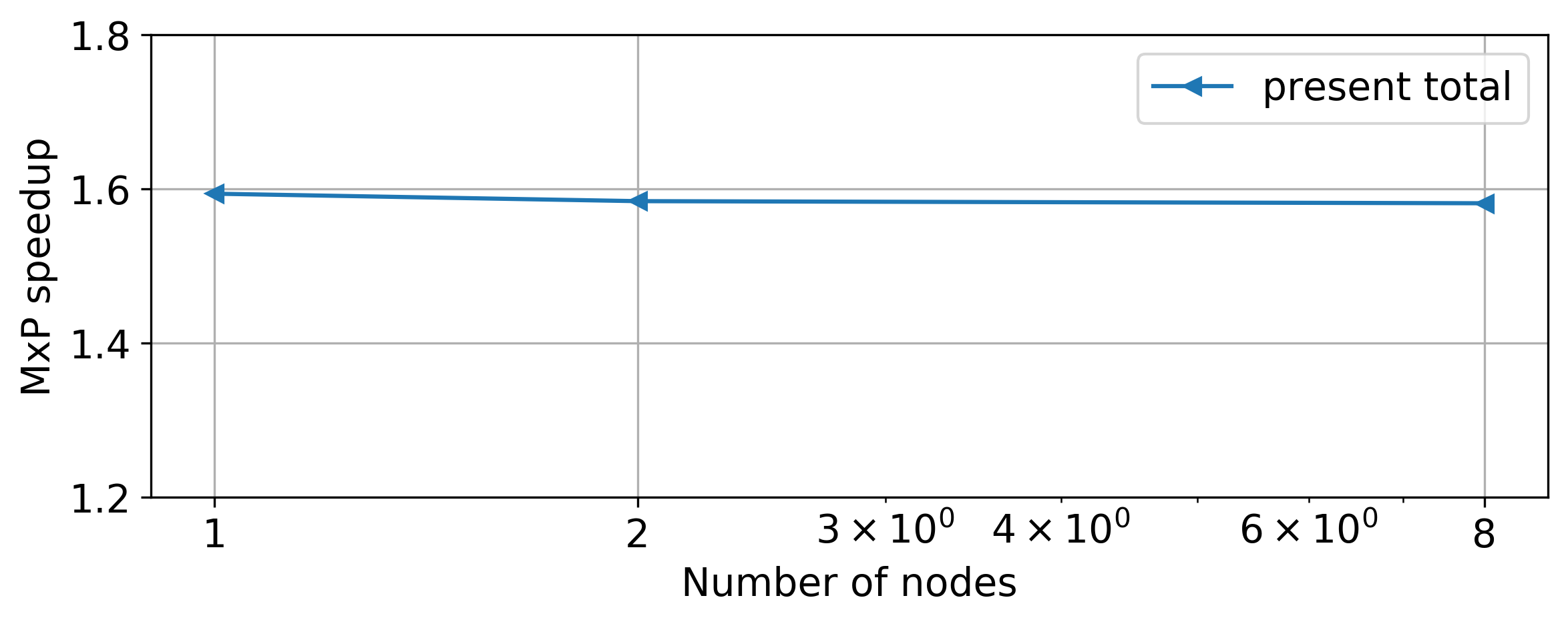}
    \end{subfigure}
    \begin{subfigure}{0.99\linewidth}
        \includegraphics[width=0.98\linewidth]{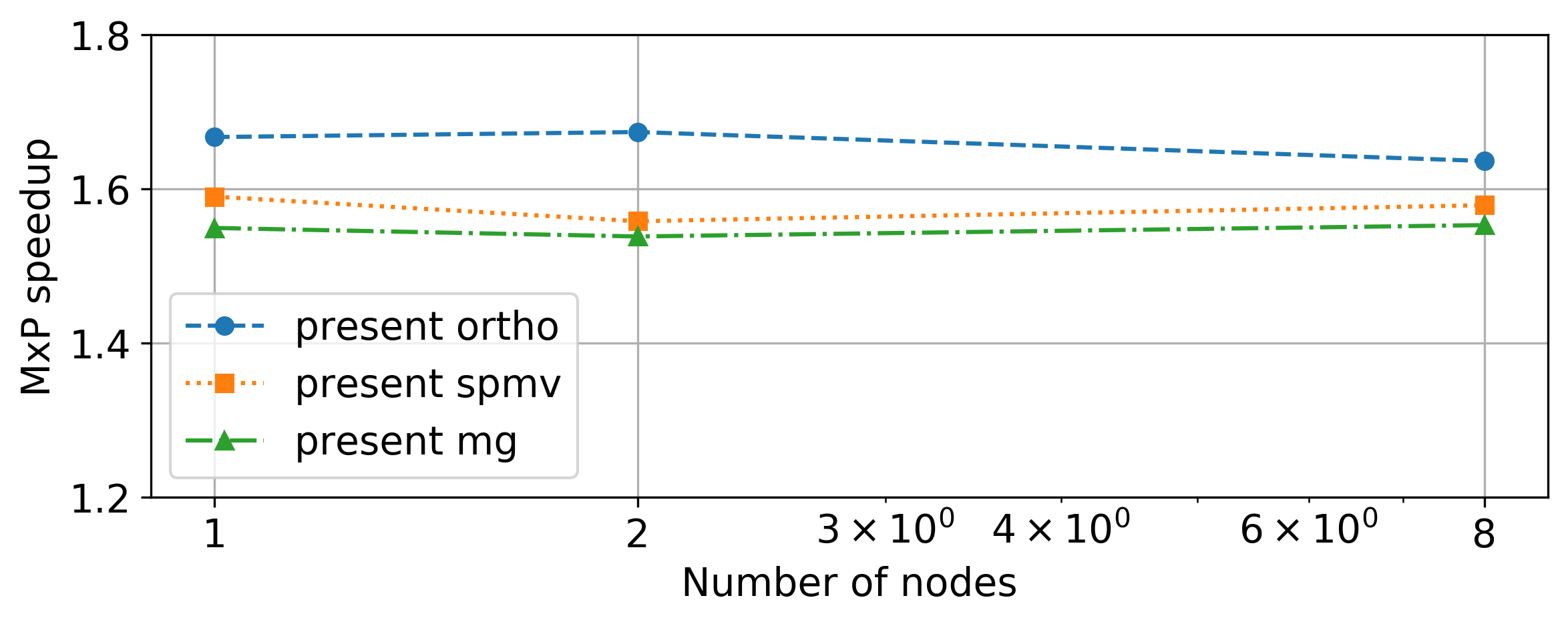}
    \end{subfigure}
    \caption{Speedups obtained on a small cluster with NVIDIA Tesla K80 GPUs}
\label{fig:andes_speedup}
\end{figure}
Because of our cross-platform implementation discussed in section \ref{sec:software}, we are able to seamlessly build and run on systems with NVIDIA GPUs. In passing, we observed (figure \ref{fig:andes_speedup}) similar speedups on a small commodity cluster containing NVIDIA K80 GPUs.

\subsection{Validation methodologies}

We compare the validation method of Yamazaki et al. \cite{hpgmp} to the new validation method described in section \ref{sec:alt_valid}.
Recall that the validation phase computes the ratio of iteration counts $\frac{n_d}{n_{ir}}$ to penalize the performance of mixed-precision GMRES-IR to include the effects of any slowdown in convergence rate.
The ratios computed by \texttt{standard} validation of Yamazaki et al. and the \texttt{fullscale} validation are shown in table \ref{tab:valid}.
As explained in section \ref{sec:hpgmp}, if the ratio is less than 1, it is considered a penalty for the mixed precision GFLOPS number.
It turns out that the \texttt{standard} small-scale validation method is more or less as stringent as \texttt{fullscale} with 10,000 iterations.
It is clear from the full-scale residual norms that, up through 8 nodes, the validation double precision solve hits the residual reduction criterion of 1e-9. However, once we get to a scale of 64 nodes, it hits the iteration limit of 10,000 iterations first, and does not reach 1e-9 residual reduction.

\begin{table}
    \begin{tabular}{|c|c|c|p{2cm}|}
    \hline
     Nodes & Std ratio & Full-scale ratio & Full-scale relative residual norm \\
    \hline
 2   & 0.968 & 0.966 & 9.98e-10 \\
 8   & 0.968 & 1.008 & 9.99e-10 \\
 64  & 0.968 & 1.050 & 1.65e-6 \\
 128 & 0.968 & 1.023 & 2.82e-6 \\
1024 & 0.968 & 1.067 & 1.154e-5 \\
4096 & 0.968 & 0.958 & 1.148e-5 \\
 \hline
    \end{tabular}
    \caption{Iteration ratios $\frac{n_d}{n_{ir}}$ for the two validation methods}
    \label{tab:valid}
    \vspace{-0.7cm}
\end{table}

\subsection{Performance analysis}

\begin{figure}
    \centering
    \includegraphics[width=0.99\linewidth]{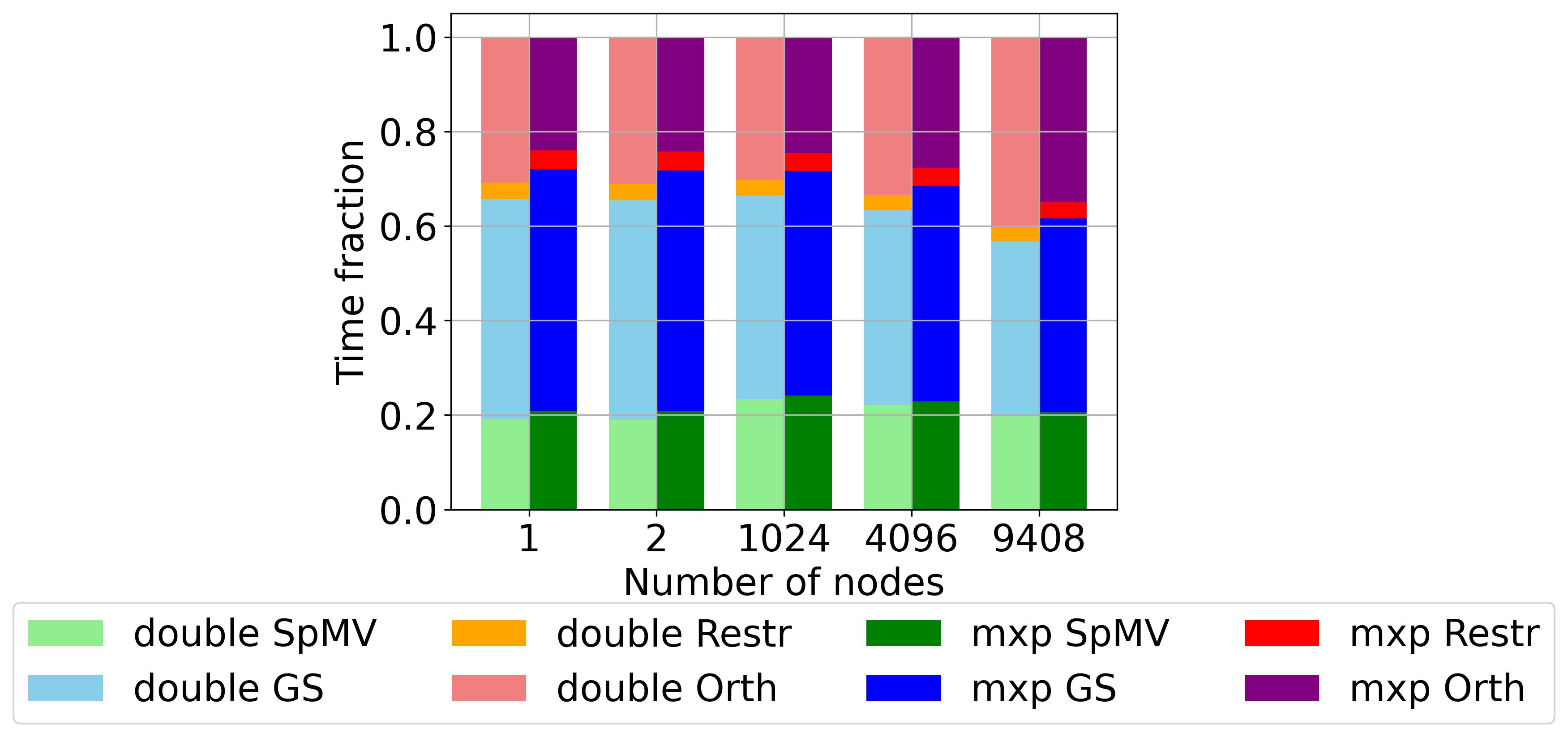}
    \caption{Breakdown of time spent in the multigrid smoother (GS), CGS2 orthogonalization (Ortho), sparse matrix vector product (SpMV) and multigrid restriction (Restr) on Frontier}
    \label{fig:stacked}
\end{figure}

We take a look at the breakdown of the time spent in the different motifs in the benchmark at two different scales in Figure \ref{fig:stacked}.
The bar chart shows the four main motifs that take nearly all the time during the mixed-precision run and the reference double-precision run of the benchmark.
As expected, the mixed-precision variant spends less time in orthogonalization, since this operation gets the most benefit from switching to single precision.
%\begin{figure}
%    \centering
%    \begin{subfigure}{0.49\linewidth}
%        \centering
%        \includegraphics[width=0.99\linewidth]{plots/pie_double_N2.png}
%        \caption{Double precision, 2 nodes}
%    \end{subfigure}
%    \begin{subfigure}{0.49\linewidth}
%        \centering
%        \includegraphics[width=0.99\linewidth]{plots/pie_double_N9408.png}
%        \caption{Double precision, 9408 nodes}
%    \end{subfigure}
%    
%    \begin{subfigure}{0.49\linewidth}
%        \centering
%        \includegraphics[width=0.99\linewidth]{plots/pie_mxp_N2.png}
%        \caption{Mixed precision, 2 nodes}
%    \end{subfigure}
%    \begin{subfigure}{0.49\linewidth}
%        \centering
%        \includegraphics[width=0.99\linewidth]{plots/pie_mxp_N9408.png}
%        \caption{Mixed precision, 9408 nodes}
%    \end{subfigure}
%    \caption{Breakdown of time spent in the multigrid smoother (GS), CGS2 orthogonalization (Ortho), sparse matrix vector product (SpMV) and multigrid restriction (Restr) on Frontier}
%    \label{fig:pies}
%\end{figure}
Going from 1 node to 9408 nodes, the full system scale, we notice that the orthogonalization takes a greater share of time, likely because the all-reduce operations in the inner product operations require more time to synchronize.

\begin{figure}[h]
    \includegraphics[width=0.98\linewidth]{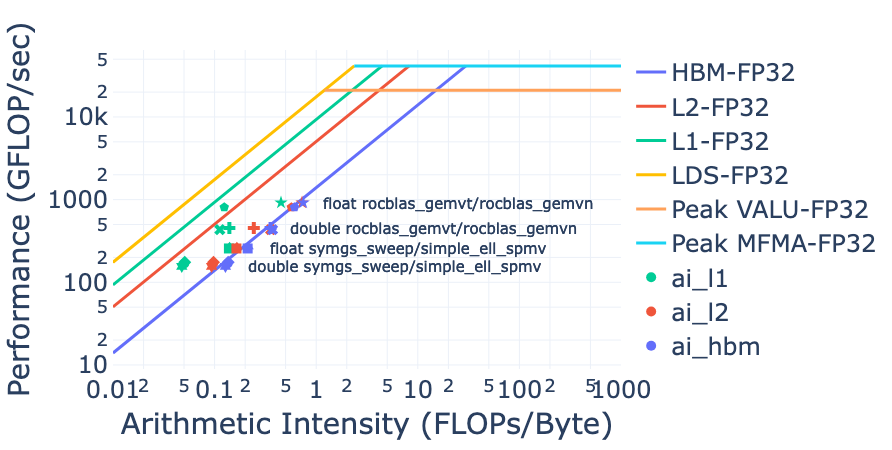}
    \caption{Roofline of the benchmark on a single GCD of an AMD MI250x. The ten most expensive kernels are depicted, eight of which are labeled. The unlabelled kernels are the double and single precision Fused SpMV-restriction, which perform similar to the Gauss-Seidel sweeps.}
    \label{fig:roofline}
\end{figure}
Figure \ref{fig:roofline} is a roofline of the most expensive kernels in the benchmark (both the optimized and reference version) on a single GCD, obtained from AMD's rocprofiler-compute.
We see that the GFLOP/s obtained for the kernels lines up at the HBM bandwidth limit. That is, despite some utilization of L2 and L1 caches, the measured throughput is at the same level as the HBM limit.
Thus, the kernels are memory bandwidth-limited, as expected. 

\begin{figure*}
    \centering
    \begin{subfigure}{0.8\linewidth}
        \centering
        \includegraphics[width=0.9\linewidth]{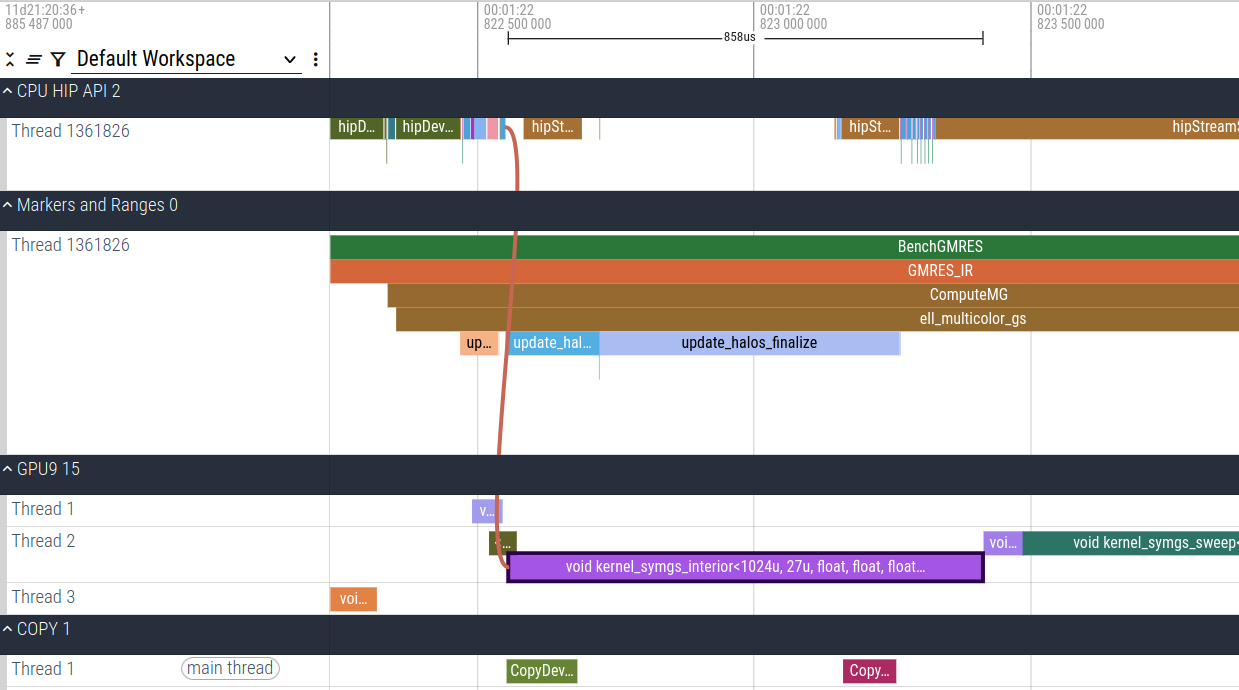}
        \caption{Fine-grid smoothing. The boxed operation is the GS kernel on the GPU.}
        \label{fig:trace_fine}
    \end{subfigure}
    
    \begin{subfigure}{0.8\linewidth}
        \centering
        \includegraphics[width=0.9\linewidth]{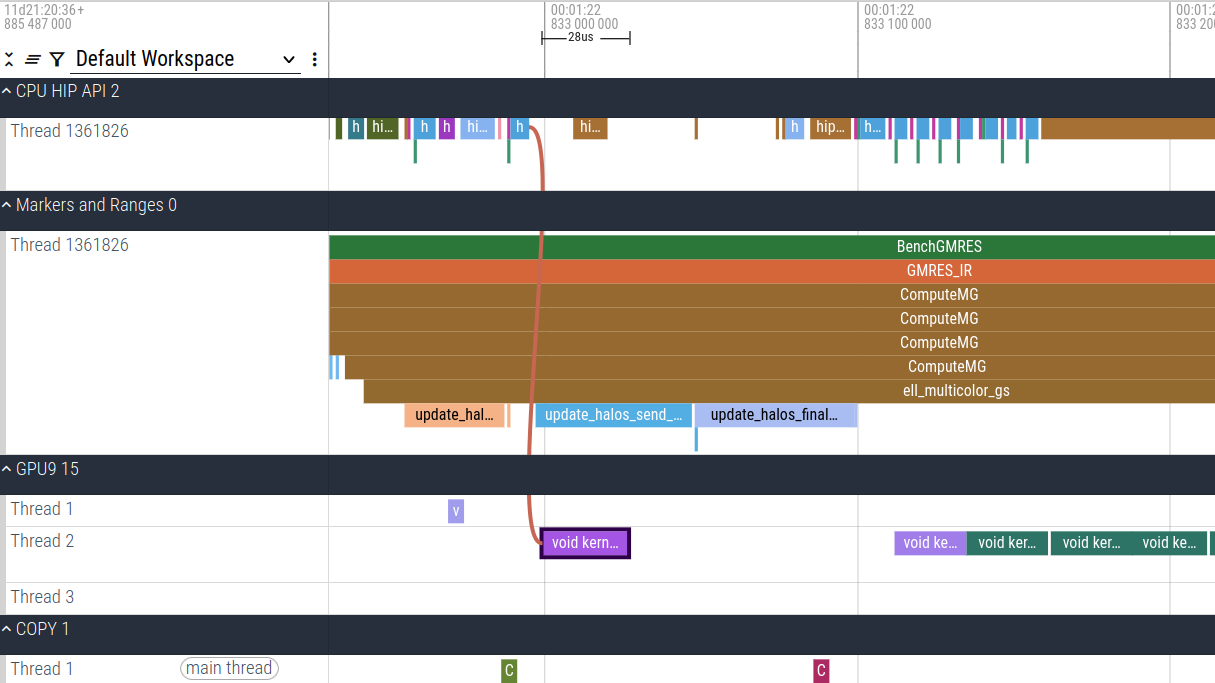}
        \caption{Coarsest grid smoothing. The boxed operation is the GS kernel on the GPU.}
        \label{fig:trace_coarse}
    \end{subfigure}
    \caption{Traces of our GMRES-IR implementation in an 8-node benchmark run on Frontier. Purple bars in the `GPU' section represent the interior Gauss-Seidel kernel, while blue bars in the `Markers and Ranges' section represent halo buffer and communications operations. In the `COPY' section, the green bar represents the device-to-host copy of the send buffer, while the red block after it is the host-to-device copy of the received data.}
    \label{fig:trace}
\end{figure*}
The compute-communication overlap achieved by our implementation can be seen on the \texttt{rocprof} traces of a `middle' rank that communicates with the maximum number of neighbors during an 8-node benchmark run on Frontier in figure \ref{fig:trace}.
On the fine grid (figure \ref{fig:trace_fine}), we can see that host-device copy after halo buffer packing, as well as the actual communications, are completely hidden by the interior Gauss-Seidel kernel on the first independent set color.
However, on the coarsest level (figure \ref{fig:trace_coarse}), only the first independent set is not sufficient to completely overlap the communication. This is because the communication surface is larger here as a ratio of the computation volume, compared to the fine grid.
Overlapping more of the Gauss-Seidel kernel with communication is possible and will be addressed in future work.
For other operations like SpMV, the halo communications are effectively hidden by interior computations on all multigrid levels.

\section{Conclusion}

Our results show that simulation workloads should seriously consider mixed-precision algorithms. 
The substantial 1.6$\times$ speedup can be obtained by carefully carrying out many of the operations in GMRES in single precision.
The fact that we obtain this speedup on well-optimized code on a wide range of scales and on more than one architecture indicates that this is a realizable speedup for production scientific applications, not an artefact of some inefficiency or a particular run configuration. 
Further, if one uses half precision strategically for parts of operations in the blue region in algorithm \ref{alg:gmres_ir_cgs2}, one can expect an even higher speedup. This will be addressed in future work.
We do accept, however, that similar to other large-scale numerical benchmarks, the matrix is artificial and the actual speedup in applications will depend on the condition numbers and pseudo-spectra of the matrices.
We introduced an option to run full-scale validation in our code and showed that the original benchmark's validation method sufficiently captures any loss of convergence rate.

Critics may argue that HPG-MxP is redundant, since HPCG already exists and both are limited by memory bandwidth. However, we argue that HPG-MxP opens up a much bigger design space by introducing mixed-precision, nonsymmetric problems and GMRES (which has different memory utilization characteristics). Being a standardized benchmark, it allows a greater variety of scientists and engineers to engage with the issue it seeks to address, and can spur innovation in achieving greater performance for PDE-based simulation workloads.

In addition, we note that the mixed-precision GMRES-IR solver requires a lower-precision copy of the system matrix. This means its overall memory utilization is more than double-precision GMRES. In order to compensate for this, we should utilize a larger mesh size while running double-precision GMRES and it can perhaps achieve a somewhat higher throughput.
The benchmark could be modified to take this into account.
In some applications, however, this may not be relevant since the matrix-free variant of GMRES \cite{chisholm_matrix_free_2009} or nonlinear GMRES \cite{washio_nka_1997} may be used. Only the low-precision matrix needs to be stored, instead of some approximate double-precision matrix, for preconditioning.

In closing, we also point out how helpful AMD's rocHPCG implementation \cite{rochpcg} has been in achieving this demonstration of HPG-MxP performance. The fact that AMD open-sourced their implementation has accelerated further progress.
In the same spirit, our code is also available open source and we provide details on building and running it in the reproducibility appendix.

\section*{Acknowledgements}
%Aditya Kashi gratefully acknowledges the useful comments from Het Mankad of the Software Engineering group at Oak Ridge National Laboratory.

%This manuscript has been authored by UT-Battelle, LLC, under contract DE-AC05-00OR22725 with the US Department of Energy (DOE). The US government retains and the publisher, by accepting the article for publication, acknowledges that the US government retains a nonexclusive, paid-up, irrevocable, worldwide license to publish or reproduce the published form of this manuscript, or allow others to do so, for US government purposes. DOE will provide public access to these results of federally sponsored research in accordance with the DOE Public Access Plan (\url{https://www.energy.gov/doe-public-access-plan}).

This research used resources of the Oak Ridge Leadership Computing Facility at the Oak Ridge National Laboratory, which is supported by the Office of Science of the U.S. Department of Energy under Contract No. DE-AC05-00OR22725.

\printbibliography

\end{document}